\documentclass[aps,prb,twocolumn, groupedaddress, showpacs]{revtex4}
\usepackage{amsmath}
\usepackage{amssymb}
\usepackage{graphicx}
\usepackage{color}
\definecolor{Blue}{rgb}{0.3,0.3,1}
\usepackage{tikz}
\usepackage{pgffor}
\usepackage{verbatim}
\begin{document}

\title{Identification of 331 quantum Hall states with Mach-Zehnder interferometry}

\author{Chenjie~Wang and D.~E.~Feldman}
\affiliation{Department of Physics, Brown University, Providence, Rhode Island 02912, USA}

\date{\today}

\begin{abstract}
It has been shown recently that non-Abelian states and the spin-polarized and unpolarized versions of the Abelian 331 state
may have identical signatures in Fabry-P\'{e}rot interferometry in the quantum Hall effect at filling factor 5/2. We calculate the Fano factor for the shot noise in a Mach-Zehnder interferometer in the 331 states and demonstrate that it differs from the Fano factor in the proposed non-Abelian states. The Fano factor depends periodically on the magnetic flux through the interferometer. Its maximal value is $2\times 1.4e$ for the 331 states with a symmetry between two flavors of quasiparticles. In the absence of such symmetry the Fano factor can reach $2\times 2.3e$. On the other hand, for the Pfaffian and anti-Pfaffian states the maximal Fano factor is $2\times 3.2e$. The period of the flux dependence of the Fano factor is one flux quantum. If only quasiparticles of one flavor can tunnel through the interferometer then the period drops to one half of the flux quantum.  We also discuss transport signatures of a general Halperin state with the filling factor $2+k/(k+2)$.
\end{abstract}


%\keywords{}

\maketitle

\begin{figure}
\centering
\includegraphics[width=2.8in]{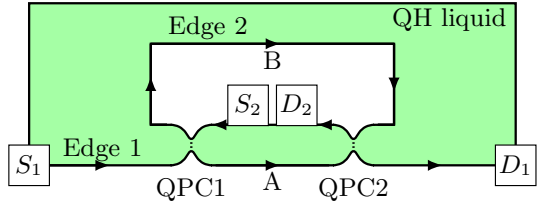}
\caption{Schematic picture of an anyonic Mach-Zehnder interferometer. Arrows indicate propagation directions of the edge modes on Edge 1 (from source $S_1$ to drain $D_1$) and Edge 2 (from source $S_2$ to drain $D_2$). Quasiparticles can tunnel between the two edges through two quantum point contacts, QPC1 and QPC2. }\label{fig1}
\end{figure}

\section{Introduction}

Two-dimensional electron systems in a strong magnetic field exhibit much beautiful physics and form many states of matter including
numerous fractional quantum Hall liquids \cite{prange87-book}.
Some of their properties such as fractional charges of elementary excitations are well understood. Less is known about the statistics of quantum Hall quasiparticles. We know that gauge invariance requires fractionally charged particles to be anyons. At the same time, a direct experimental observation of anyonic statistics poses a major challenge. This challenge has recently attracted much attention because of a possibility of non-Abelian anyonic statistics at some quantum Hall filling factors \cite{nayak08}. In contrast to ``tamer'' Abelian particles, non-Abelian anyons change their quantum state after one particle makes slowly a full circle around other anyons. This property can be used for topological quantum computation \cite{kitaev03}.
Possible application as well as intrinsic interest of such unusual particles have stimulated attempts to find non-Abelian anyons in nature. In particular, a possibility of non-Abelian statistics was predicted at filling factors $5/2$ and $7/2$ \cite{overbosch08,moore91,lee07,levin07}. However, the nature of the quantum Hall states at those filling factors remains an open question with theoretical proposals including both Abelian
and non-Abelian states \cite{overbosch08,halperin83,wenzee92,yss92,blockwen90}.

Numerical simulations \cite{Morf98,RH00,MS08,FRYND09} with small systems of 8-20 electrons provide support to non-Abelian Pfaffian and anti-Pffafian states. This support is however not unanimous, see, e.g., Ref.~\onlinecite{TRJ07}. At the same time, recent
experiments \cite{Dean08,Zhang10,MStern10}
suggest an unpolarized state at $\nu=5/2$. Zero spin polarization is incompatible with either Pfaffian or anti-Pfaffian states. Some of the experiments can be understood in terms of disorder-generated skyrmions
\cite{WMSC10}
in a Pfaffian state. However,
such an explanation does not apply to the most recent optical experiment \cite{MStern10}
and the simplest interpretation of the existing limited experimental data is in terms
of zero polarization \cite{DGZ10,Jain10}. The simplest unpolarized state is the Halperin 331 state
\cite{overbosch08}. Note that the results of the tunneling experiment \cite{radu08}
are compatible with both 331 and anti-Pfaffian states.
 In a closely related problem of quantum Hall bilayers at filling factor
5/2, numerics supports the existence of both 331 and Pfaffian states, separated by a phase transition \cite{PD10}. Thus, it is important to find a way to identify and distinguish from each other the 331, Pfaffian and anti-Pfaffian states.

The key difference lies in non-Abelian quasiparticle statistics of the Pfaffian and anti-Pfaffian states
versus Abelian statistics in the 331 state.
In order to determine the statistics of anyons at $\nu=5/2$ and 7/2 several experiments were proposed and some of them have been or are being currently implemented (for a review, see Ref.~\onlinecite{stern10}). Despite those efforts the statistics in the 5/2- and 7/2-states remains an open question. One of the issues concerns ambiguities in the interpretation of the experimental data.
In particular, the most elegant and conceptually simple approach to detecting non-Abelian anyons is based on Fabry-P\'{e}rot interferometry \cite{fradkin98,dassarma05,stern06,bonderson06,willet09}.
It was found recently \cite{bonderson09,stern09,ilan10}
that a Fabry-P\'{e}rot interferometer may produce identical interference and Coulomb blockade patterns in transport experiments
with the non-Abelian Pfaffian and anti-Pfaffian states and
the spin-polarized \cite{overbosch08,wenzee92,yss92,blockwen90} and unpolarized\cite{halperin83} versions of the Abelian 331 state.
Similarity between the patterns is only present in the case of the exact or approximate symmetry between two quasiparticle flavors in the 331 states and one may hope to remove such symmetry by some perturbation or a change in  the conditions of the experiment \cite{stern09}. However, in the absence of an established theory of 5/2- and 7/2-states, it is hard to tell whether the flavor symmetry is or is not present and if a particular perturbation would make it possible to distinguish Abelian and non-Abelian states. Thus, it is desirable to have another approach to interferometry such that Abelian states would never mimic non-Abelian.

In this paper we show that in an anyonic Mach-Zehnder interferometer \cite{ji03,law06,feldman06,feldman07,law08, ponomarenko09},
signatures are different for the Pfaffian and anti-Pfaffian states on the one hand and the polarized and unpolarized 331 states on the other hand. This conclusion is true both in the presence and absence of the flavor symmetry. We calculate the low-temperature zero-frequency noise in the interferometer in the limit of weak tunneling through the device. In such case the noise is related to the current as $S=2e^*I$, where $2e^*$ is the Fano factor. The Fano factor exhibits a periodic dependence on the magnetic flux. Its maximal value as a function of the flux was calculated for the Pfaffian and anti-Pfaffian states in Ref.~\onlinecite{feldman07} and equals $2e^*=2\times 3.2e$, where $e$ is an electron charge. We show that in the 331 states with flavor symmetry the maximal value of the Fano factor is $2e^*=2\times 1.4e$.
In the absence of the symmetry the Fano factor can reach the maximal value of $2\times 2.3e$ in the 331 states. Thus, the maximal Fano factor gives an unambiguous way to distinguish
the Abelian 331 quantum Hall liquids from the proposed non-Abelian states. We also calculate the electric current through the interferometer as a function of the flux and voltage. The results for the 331 state differ from the case of the Pfaffian state but the difference between the two $I$-$V$ curves is small.

The paper is organized as follows. First, we briefly discuss the 331 states. Next, we review the structure of the anyonic Mach-Zehnder interferometer
in Sec.~\ref{sec:MZI}.
We calculate the zero-temperature current in Sec.~\ref{sec:current}
and zero-temperature zero-frequency shot noise in Sec.~\ref{sec:noise}. We summarize our results in Sec.~\ref{sec:summary}. The Appendix contains a detailed discussion of general
Halperin states at filling factor $\nu=2+k/(k+2)$.

\section{Statistics in the 331 state}
\label{sec:statistics}

A review of different proposed states for the filling factor $5/2$ can be found in Ref.~\onlinecite{overbosch08}.
Here we summarize the properties of the 331 states \cite{halperin83,wenzee92,yss92,blockwen90}.

Two different variants of the 331 state are described in the literature. The spin-unpolarized version was introduced in Ref.~\onlinecite{halperin83}. It can be understood as a bilayer state
with two spin components playing the role of the layers.
The filling factor in each layer is $1/4$. The spin-polarized version \cite{overbosch08,wenzee92,yss92,blockwen90}
can be described in terms of the condensation
of the charge-$2e/3$ quasiparticles on top of the Laughlin $\nu=1/3$ state.
The two states differ in many respects but have the same key features: the topological order and the statistics of quasiparticles \cite{overbosch08}.
Since anyonic interferometry is only sensitive to the quasiparticle statistics, it cannot distinguish the two states. Below we describe
the statistics with the help of the $K$-matrix formalism, Ref.~\onlinecite{wen04-book}.
We only focus on the half-filled Landau level. Integer edge channels formed
in the lower completely filled Landau levels are unimportant for our problem since the transport through
the interferometer is dominated by the tunneling of
fractionally charged excitations on top of the half-filled Landau level.

The $K$-matrix formalism encodes the information about quasiparticles in terms of a matrix $K$ and a charge vector ${\bf t}$.
 Each elementary
excitation is described by a vector ${\bf l}_n$ with integer components.
In the 331 states all edge modes propagate in the same direction. In such case the scaling dimensions of quasiparticle creation and annihilation
operators are independent of the interactions between the modes and are given by $h_n={\bf l}_n K^{-1}{\bf l}_n^T$. We are only interested in
the most relevant quasiparticle operators.
The quasiparticle charge $Q=-e{\bf t}_n K^{-1}{\bf l}_n^T$. Transport through the Mach-Zehnder interferometer depends on statistical phases
accumulated by the wave function when one particle makes a full circle around another.
The phase accumulated by particle 1 moving around particle 2 equals
$\theta_{12}=2\pi{\bf l}_1 K^{-1}{\bf l}_2^T$.

The spin-unpolarized 331 state\cite{halperin83} can be described as a bilayer state with $\nu=\frac{1}{4}$ in each layer.
The $K$-matrix is
\begin{equation}
K = \left(
\begin{array}{cc}
3& 1\\
1& 3
\end{array}\right).
\end{equation}
The charge vector {\bf t}=(1,1).
The two most relevant quasiparticles are characterized by the $l$-vectors ${\bf l}_1=(1,0)$ and ${\bf l}_2=(0,1)$.
Both particles have charge $e/4$. This elementary excitation charge agrees with experiments\cite{dolev08,radu08,willet09}.
When a particle makes a circle around an identical particle it accumulates the phase
$\phi_{11}=\phi_{22}=3\pi/4$. The mutual statistical phase of two different particles is $\phi_{12}=-\pi/4$.
We will also need to know what phases are accumulated when a charge-$e/4$ quasiparticle $q_0$ moves around a composite
anyon built from several  $e/4$-quasiparticles $q_1,\ldots,q_k$. In the Abelian 331 state such statistical phase is simply
the sum of mutual statistical phases of $q_0$ with each of the $q_m$ particles.

The spin-polarized 331 state\cite{overbosch08} is formed by the condensation of the charge-$2e/3$ quasiparticles
on top of the Laughlin $\nu=1/3$ state. This state is characterized by the $K$-matrix
\begin{equation}
K = \left(
\begin{array}{cc}
3& -2\\
-2& 4
\end{array}\right).
\end{equation}
and the charge vector ${\bf t}=(1,0)$. Calculations of the electric charges and statistical phases are the same as above.
The two most relevant elementary excitations carry charges $e/4$ again and are characterized by vectors ${\bf l}_1=(0,1)$ and ${\bf l}_2=(1,-1)$.
All statistical phases are the same as in the spin-unpolarized 331 state.

In fact, the two versions of the 331 state are topologically equivalent\cite{wen04-book}, since the two $K$-matrices satisfy
\begin{equation}
\left(
\begin{array}{cc}
3& 1\\
1& 3
\end{array}\right)= W
\left(
\begin{array}{cc}
3& -2\\
-2& 4
\end{array}\right)W^{T}, \quad
W=\left(
\begin{array}{cc}
1& 1\\
1& 0
\end{array}\right)
\end{equation}
and the charge vectors satisfy
\begin{equation}
(1,1)=(1,0) W^T
\end{equation}
This explains why the two states have the same quasiparticle charges and braiding statistics.

\section{Mach-Zehnder interferometer}
\label{sec:MZI}

Figure~\ref{fig1} shows a sketch of the Mach-Zehnder geometry. Because of the bulk energy gap, the low-energy physics is determined by chiral edge modes.
Charge flows along Edge 1 from source $S_1$ to drain $D_1$ and along Edge 2 from source $S_2$ to drain $D_2$. Quasiparticles tunnel
between the two edges at
quantum point contacts QPC1 and QPC2. If one keeps $S_1$ at a positive voltage $V$ and the other source and drains are grounded then there is
a net quasiparticle flow into Edge~2 and a net tunneling current. The current is measured at drain $D_2$.

As discussed above, there are two flavors of charge-$e/4$ quasiparticles in the 331 state.
Let us denote their topological charges (or flavors) as $a$ and $b$.
Since the most relevant quasiparticle operators create particles of these two types,
we consider only the tunneling of  $e/4$-quasiparticles with flavors $a$ and $b$ below.
We focus on the limit of small tunneling amplitudes between the edges.
In such case the problem can be accessed with perturbation theory.
We denote the small tunneling amplitudes  at the two point contacts as $\Gamma_k^x$,
where $k=1,2$ is the number of the point contact and $x=a,b$ is the topological
charge of the tunneling quasiparticle. The tunneling rate  from Edge 1 to Edge 2 can be found from the Fermi golden rule.
It depends on the tunneling amplitudes and on the phase difference for the quasiparticles which follow from S1 to D2 through QPC1 and QPC2.
The latter consists of two contributions: the Aharonov-Bohm phase due to the external magnetic field and the statistical
phase accumulated by a quasiparticle
making a full circle around the ``hole'' in the interferometer.
The statistical phase is determined by the total topological charge that tunneled previously
between the edges. Indeed, the total topological charge of Edge 2 can only change
during tunneling events at QPC1 and QPC2. Charge exchange with the Fermi-liquid
drain D2 and source S2 cannot affect the topological charge of the edge.
Taking into account that the edges are chiral we see that all topological charge that
previously tunneled into Edge 2 accumulates inside the loop QPC1-A-QPC2-B-QPC1.
We will denote that loop as $\mathfrak L$ below.
 Certainly, the accumulated topological charge only assumes a discrete set
of values and hence changes quasiperiodically as a
function of time. We find the following tunneling rate from edge 1 to edge 2 for a quasiparticle of flavor $x$
(cf. Ref.~\onlinecite{feldman07})
\begin{equation}
w_{x,d}^+ = r_1(|\Gamma_1^x|^2 + |\Gamma_2^x |^2) + (r_2\Gamma_1^x\Gamma_2^{x*} e^{i\phi_{\rm mag}+i\phi_{xd}}+ {\rm c.c.}),\label{eq1}
\end{equation}
where $r_1(V,T,x)$ and $r_2(V,T,x)$ depend on the voltage, temperature and quasiparticle flavor, $d$
is the topological charge trapped in the interferometer before the tunneling event,
$\phi_{xd}$ the statistical phase discussed in the previous section, and the Aharonov-Bohm phase
$\phi_{\rm mag}=\pi\Phi/(2\Phi_0)$ is expressed in terms of the magnetic flux $\Phi$ through the loop $\mathfrak L$ and the flux quantum $\Phi_0=hc/e$.
$r_{1,2}$ cannot be calculated without a detailed understanding of the edge physics. Fortunately, we will not need
 such a calculation to determine the main features of the current and noise in the interferometer.
Eq.~(\ref{eq1}) assumes that the two tunneling amplitudes are small and hence the mean time between two consecutive tunneling events is much longer than
the time spent by a tunneling quasiparticle between the point contacts.
At a nonzero temperature, quasiparticles are allowed to tunnel from Edge 2 to Edge 1 which has a higher potential.
The corresponding tunneling rate $w_{x,d+x}^-=\exp[-eV/(4k_BT)]w_{x,d}^+$ is connected to $w^+$, Eq.~(\ref{eq1}), by the detailed balance principle. Here $d+x$ is the topological charge of the fusion of anyons with topological charges $d$ and $x$, i.e., the topological charge of Edge 2 before the tunneling event. In what follows we
concentrate on the limit of low temperatures and neglect $w^-$.

Our main focus will be on the situation with the exact or approximate flavor symmetry,
i.e., we assume that the tunneling amplitudes  $\Gamma_k^x$ and coefficients $r_k(V,T,x)$ in (\ref{eq1}) do
not depend on the flavor $x$. Since $\Gamma$'s and $r$'s only enter the transition rates in the combinations
$r_1(V,T,x)|\Gamma_{1,2}^x|^2$ and $r_2(V,T,x)\Gamma_1^x\Gamma_2^{x*}$, the results do not change if the tunneling amplitudes depend on the flavor but
the above combination are the same for $x=a$ and $b$. As discussed in the introduction, Fabry-P\'{e}rot interferometry cannot distinguish the 331 states
with flavor symmetry from non-Abelian states. We will see below that no such issue exists for Mach-Zehnder interferometry. We will also briefly discuss
signatures of the 331 states without flavor symmetry in the shot noise in a Mach-Zehnder interferometer.

\begin{figure*}
\centering
\includegraphics[width=5in]{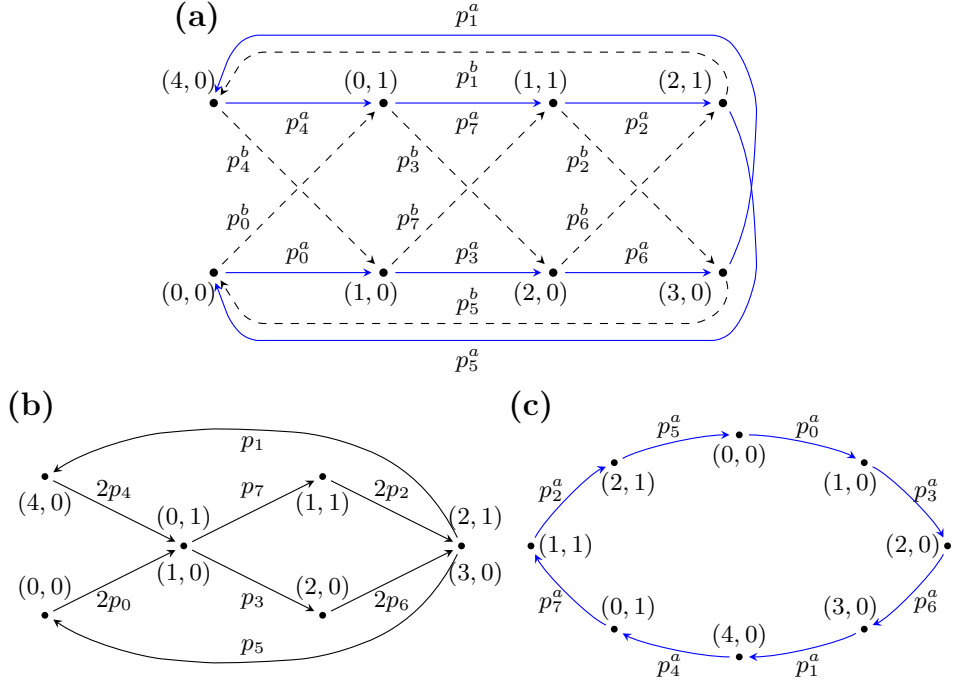}
\caption{Possible states of a Mach-Zehnder interferometer in  the $331$ state. Panel (a) shows a general case, eight possible states labeled by topological charges and the transition rates between them. Arrows show the allowed transitions at zero temperature. Solid blue lines represent tunneling events involving quasiparticles of flavor $a$, and dashed black lines
represent tunneling events involving particles of flavor $b$.
Special cases with $p_k^a=p_k^b\equiv p_k$ and $p_k^b=0$ are illustrated in Panels (b) and (c) respectively.}
\label{fig2}
\end{figure*}

\section{Electric current}

\label{sec:current}
We are now in the position to calculate the tunneling current $I$. From now on, we will only use the language of the spin-unpolarized 331 state
in which flavor $a$  can be understood as the $l$-vector ${\bf a}=(1,0)$ and $b$ as ${\bf b}=(0,1)$.
From Sec.~\ref{sec:statistics} we know how to evaluate the statistical phase in (\ref{eq1}).
The tunneling rate depends on the topological charge ${\bf d}=(m,n)$ trapped in the interferometer.
Each tunneling event of a particle with flavor $a$ changes ${\bf d}\rightarrow(m+1,n)$. If a particle of flavor $b$ tunnels then ${\bf d}\rightarrow(m,n+1)$.

Any anyon has a trivial mutual statistical phase $2\pi k$ with an electron \cite{kitaev-review}.
Combining this condition with the knowledge of the electron charge one finds
that the $l$-vectors $(1,3)$ and $(3,1)$ describe electrons.
Thus, the topological charges ${\bf d}$ and ${\bf d}'={\bf d}+n_1(1,3)+n_2(3,1)$ can be viewed as identical
since
anyons accumulate identical topological phases moving around charges ${\bf d}$ and ${\bf d}'$.
One can easily see that in a 331 state, the trapped topological charge falls into one of eight equivalence classes which mark eight
 possible topological states of the area enclosed by the loop $\mathfrak L$.
Fig.~\ref{fig2}(a) shows the 8 states and possible transitions between them
due to anyon tunneling between the edges at zero temperature. The transition rates shown in Fig.~\ref{fig2} are given by the equation
\begin{equation}
p_k^x = A^x[1+u^x\cos(\pi\Phi/(2\Phi_0)+\pi k/4 + \delta^x)]
\label{eqp}
\end{equation}
where $x=a$ or $b$, and $k=0,1,\ldots,7$. The parameters $A^x=r_1(|\Gamma_1^x|^2+|\Gamma_2^x|^2)$,
$u^x=2|r_2\Gamma_1^x\Gamma_2^x|/[r_1(|\Gamma_1^x|^2+|\Gamma_2^x|^2)]$ and $\delta^x= {\rm arg}(r_2\Gamma_1^x\Gamma_2^{x*}) $.
In the presence of the flavor symmetry the diagram in Fig.~\ref{fig2}(a) simplifies to Fig.~\ref{fig2}(b) with six instead of eight vertexes.
Two pairs of vertexes in Fig.~\ref{fig2} (a) are merged into single vertexes in Fig.~\ref{fig2} (b).
This is legitimate due to numerous equalities between tunneling rates.
In Fig.~\ref{fig2} (b), we use the notation $p_k=p_k^a=p_k^b$. We will also denote $A=A^a=A^b$, $u=u^a=u^b$ and $\delta=\delta^a=\delta^b$
in the presence of the flavor symmetry. Fig.~\ref{fig2}(c) illustrates another simple limit in which $p_k^b =0$.

\begin{figure}[t]
\centering
\includegraphics[width=3in]{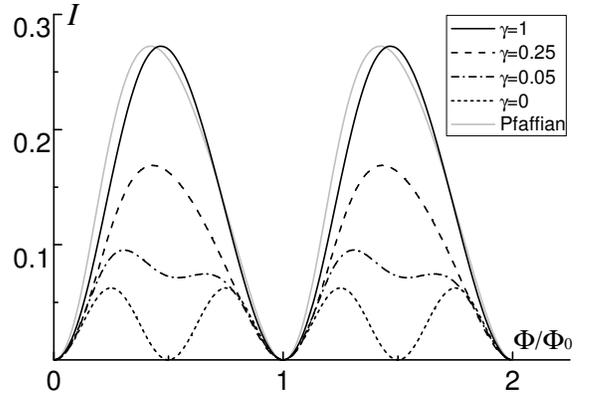}
\caption{Flux-dependence of the tunneling current in the 331 and Pfaffian states. We set $A^a=1$, $u^a=u^b=1$, and $\delta^a=\delta^b=0$ for all curves
for the 331 state. Different curves correspond to different values of $\gamma=A^b/A^a$ in the 331 state. The curve for the Pfaffian state is plotted according to Eq.~(8) in Ref.~\onlinecite{feldman06} with $r_{11}^+=r_{12}^+=1$ and $\Gamma_1=\Gamma_2$ such that the maximum matches the maximum of the curve for the 331 state with $\gamma=1$.}\label{fig3}
\end{figure}

The transition rates can be used to write down a kinetic equation for the probabilities $f_d$ of different topological charges
$d$ trapped in the interferometer. In terms of the distribution function $f_d$ the average tunneling current between the edges
\begin{equation}
I= \frac{e}{4} \sum_{d} f_d (w_{a,d}^+ + w_{b,d}^+ - w_{a,d}^- - w_{b,d}^-), \label{eq:current}
\end{equation}
where $d$ goes over the eight possible states in Fig.~\ref{fig2}(a). The distribution function satisfies the steady state equation
\begin{align}
0=\frac{d f_d}{dt}= & \sum_{x=a,b}(f_{d-x}w_{x,d-x}^+ + f_{d+x}w_{x,d+x}^-) \nonumber\\
& - \sum_{x=a,b}f_{d}(w_{x,d}^+ + w_{x,d}^-),\label{eq:steady}
\end{align}
and the normalization condition $\sum_d f_d=1$.
In Eq.~(\ref{eq:steady}), $(d+x)$ means the topological charge of the fusion of anyons with topological charges $d$ and $x$. The fusion of $x$ and $d-x$
has topological charge $d$.
In the absence of the flavor symmetry, the tunneling current can be found with a lengthy but straightforward calculation from Eqs. (\ref{eq:current})
and (\ref{eq:steady}).
In the presence of the flavor symmetry, the low-temperature current can be easily calculated from Eqs. (\ref{eq:current},\ref{eq:steady})
in the picture with six distinct topological charges, Fig.~\ref{fig2} (b). One obtains the following result
\begin{widetext}
\begin{equation}
\label{eq:current2}
I = \frac{eA}{2} \frac{1-u^2+\frac{u^4}{8}(1-\cos(2\pi\Phi/\Phi_0 +4\delta))}{1-(\frac{3}{4}-\frac{1}{4\sqrt{2}})u^2+\frac{u^4}{16} \left[(1-\frac{1}{\sqrt2})(1-\cos(2\pi\Phi/\Phi_0+4\delta))-\frac{1}{\sqrt2}\sin(2\pi\Phi/\Phi_0+4\delta)\right]}
\end{equation}
\end{widetext}
This formula is quite similar to the expression for the current in the Pfaffian state (cf. Eq.~(8) in Ref.~\onlinecite{feldman06}).
The similarity originates from the similarity of the diagram Fig.\ref{fig2}(b) with a corresponding diagram in the Pfaffian state
\cite{feldman06}.
They have the same topology with six vertexes including two ``cross-roads''.
Still, in contrast to the Fabry-P\'{e}rot case, the expressions for the current
are not identical for the 331 and Pfaffian states. Similar to the Pfaffian state \cite{feldman06}, the expression for the current
for the opposite voltage sign can be obtained
by changing both the overall sign of the current and the sign before $\sin(2\pi \Phi/\Phi_0+4\delta)$ in the denominator. Thus, the $I$-$V$ curve is
asymmetric just like in the Pfaffian case.
The current depends periodically on the magnetic flux with the period $\Phi_0$ in accordance with the Byers-Yang theorem \cite{BYt}. We do not include an analytical expression for the current in the absence of the flavor symmetry since it is lengthy. $I(\Phi)$ is plotted in Fig.~3 for the Pfaffian and 331 states for different values of
$\gamma=A^b/A^a$ at $u=1$ which maximizes the visibility of the Aharonov-Bohm oscillations.

Another simple limit corresponds to the situation with $p_k^b=0$ (Fig.~\ref{fig2}(c)). In that case, $I(\Phi)$ has a reduced period $\Phi_0/2$.
The period reduction can be understood from the diagram Fig.~\ref{fig2}(c). The system can return to its initial state only after eight tunneling events instead of four in
Fig.~\ref{fig2} (b). This means a transfer of  the charge $2e$ in each cycle of tunneling events.
Such ``Cooper pair'' charge agrees with the ``superconductor'' periodicity.

\section{Shot noise}
\label{sec:noise}

Shot noise measurements have helped to determine the charges of elementary excitations at several quantum Hall filling factors including 5/2
\cite{dolev08,noise1,noise2}.
In the Mach-Zehnder geometry, zero-frequency noise contains also information about the quasiparticle statistics. Below we calculate
the noise in the 331 states in the limit of weak tunneling. We assume that the temperature is much lower than the voltage bias. In such case
it is possible to neglect the temperature and perform calculations at $T=0$.

Shot noise is defined as the Fourier transform of the current-current correlation function,
\begin{equation}
S(\omega) = \int_{-\infty}^{+\infty}\langle\hat I(0)\hat I(t) + \hat I(t) \hat I(0)\rangle \exp(i\omega t)dt.
\end{equation}
Below we consider $\omega=0$ only. In the weak tunneling limit, the noise can be expressed as $S=2e^*I$, where $2e^*$ is known as the Fano factor.
$e^*$ can be understood as an effective charge tunneling through the interferometer. We will see that the Fano factor is different
in the Pfaffian state and the 331 states with or without the flavor symmetry.
In the Mach-Zehnder interferometer, the Fano factor $2e^*(\Phi)$  exhibits oscillations as a function of the magnetic flux.
For Laughlin states, the maximal $e^*$ can never exceed $1.0e$, while in the Pfaffian state $e^*$ can be as large as $3.2e$, Ref.~\onlinecite{feldman07}.
We will see that the maximal Fano factor is lower in the 331 state than in the Pfaffian state.

\begin{figure}[t]
\centering
\includegraphics[width=3in]{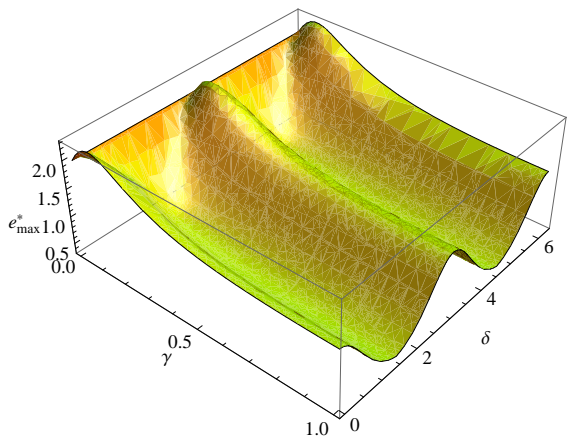}
\caption{The maximal $e^*$ as a function of $\gamma$ and $\delta$.}\label{fig3D}
\end{figure}

A zero-frequency shot noise can be conveniently connected with the fluctuation of the charge transmitted through the interferometer
over a long time $t$ (cf. Ref.~\onlinecite{feldman07}):
\begin{equation}
S/2 = \lim_{t\rightarrow \infty} \langle \delta Q^2(t)\rangle /t,
\end{equation}
where $\delta Q(t)$ is the fluctuation of the charge $Q(t)$ that has tunneled into the loop $\mathfrak L$ during the time $t$.
The average current $I = \lim _{t\rightarrow \infty} \langle Q(t)\rangle/t$.
We will use the generating function method developed in Ref.~\onlinecite{feldman07} (see also Ref.~\onlinecite{law08}) to calculate the zero-frequency shot noise and the Fano factor.
Suppose for certainty that the initial state inside the loop $\mathfrak L$ has topological charge $(0,0)$. Our results will not depend on this choice
of the initial topological charge.  The electric charge that had
tunneled through the interferometer, $Q(t)$, is zero at $t=0$, $Q(0)=0$.
The system evolves over a time period $t$ according to a kinetic equation with the transition rates from the diagram Fig.~\ref{fig2}(a).
The probability of finding the interferometer in the state with the topological charge $d$ on Edge 2
and the transmitted electric charge $Q(t)=k\frac{e}{4}$
will be denoted as $f_{d,k}(t)$ below.
For a given $d$, only certain $k$'s are possible (e.g., if $d=(0,1)$ then $k$ can only be $4n+1$ with an integer $n$).
The distribution function satisfies the kinetic equation
\begin{align}
\frac{d f_{d,k}(t)}{dt}= & \sum_{x=a,b}(f_{d-x,k-1}w_{x,d-x}^+ + f_{d+x, k+1}w_{x, d+x}^-)\nonumber \\ & - \sum_{x=a,b}f_{d,k}(w_{x,d}^++ w_{x,d}^-).\label{eq:distrib}
\end{align}
Let us now define the generating function
\begin{equation}
f_d(s,t)=\sum_{k} f_{d,k}(t)s^{k},
\end{equation}
where the summation extends over all possible $k$'s for a given $d$. One can express the average transmitted charge $\langle Q(t)\rangle =e\sum_k kf_{d,k}(t)/4$ and its fluctuation as
\begin{equation}
\langle Q(t)\rangle  = \frac{e}{4}\left.\left( \frac{d}{ds} \sum_{d}f_d(s,t)\right)\right|_{s=1} \label{eq:Q1}
\end{equation}
and
\begin{equation}
\langle \delta Q^2(t) \rangle = \left(\frac{e}{4}\right)^2 \left(\frac{d}{ds}s\frac{d}{ds}\left.\sum_{d}f_d(s,t)\right)\right|_{s=1} -\langle Q(t)\rangle^2, \label{eq:Q2}
\end{equation}
where the angular brackets stay for the average with respect to the distribution function $f_{d,k}(t)$.
It is easy to see that $f_d(1,t)$ equals the probability $f_d(t)$ introduced in Sec.~\ref{sec:current}. The kinetic equation (\ref{eq:distrib}) then reduces to
\begin{align}
\frac{d}{dt}f_d(s,t) =&  \sum_{x=a,b} (sf_{d-x}w_{x,d-x}^+ + \frac{1}{s} f_{d+x} w_{x,d+x}^-)\nonumber \\ &- \sum_{x=a,b} f_{d} (w_{x,d}^++ w_{x,d}^-) \label{eq:kinetic}
\end{align}
The above equation can be rewritten in a matrix form, $\dot{\vec{f}}(s,t) = M(s)\vec f(s,t)$, where $M(s)$ is a time-independent $8\times8$ matrix.
We can solve the above linear equation and obtain
\begin{equation}
f_{d}(s,t)=\sum_{i=0}^7\xi_{d,i}(s)e^{\lambda_i(s)t}P_i(t),\label{eq:solution}
\end{equation}
where $\lambda_i(s)$ are eigenvalues of $M(s)$, $\xi_{d,i}$ constants and $P_i(t)$ polynomials, $P_i=1$ for nondegenerate eigenvalues $\lambda_i$.
The solution can be further simplified with the help of the Rohrbach theorem \cite{rohrbach}. Indeed,
from (\ref{eq:kinetic}) we see that $M(s)$ has negative diagonal elements, non-negative off-diagonal elements,
and the sum of the elements of each column is zero at $s=1$. In such situation the Rohrbach theorem applies.
According to the theorem, the eigenvalue of $M(s=1)$ with the maximal real part
is nondegenerate and equals zero.
All other eigenvalues have negative real parts.
For $s$ close to 1, the maximal eigenvalue $\lambda_0(s)$ should also be nondegenerate by continuity of the eigenvalues as functions of
$s$. Thus, the $\lambda_0(s)$ term in Eq.~(\ref{eq:solution}) dominates  at large time $t$ and $P_0(t)=1$.

According to (\ref{eq:Q1}), (\ref{eq:Q2}) and (\ref{eq:solution}) together with the conservation of probability $\sum_{d} f_d(1,t)=1$, we have
\begin{align}
I&=\frac{e}{4}\lambda_0'(1)\nonumber\\
S&=(e/4)^2\left[\lambda_0''(1)+\lambda_0'(1)\right] \nonumber\\
e^* & = \frac{e}{4} [1+\lambda_0''(1)/\lambda_0'(1)]
\end{align}
The derivatives of $\lambda_0(s)$ can be evaluated in the following way. Define a function
\begin{equation}
G(s,\lambda)=\det(M(s)-\lambda E)
\end{equation}
where $\det$ is the determinant of a matrix and $E$ is the identity matrix. We know that $\lambda_0(s)$ satisfies the equation $G(s,\lambda_0(s))=0$.
Then by differentiating this equation one gets
\begin{align}
\lambda_0'(1) &= -G_{s}/G_{\lambda}|_{(s,\lambda)=(1,0)}\nonumber\\
\lambda_0''(1) &= -(G_{ss}G_{\lambda}^2+G_{\lambda\lambda}G_{s}^2-2G_{s\lambda}G_{s}G_{\lambda})/G_{\lambda}^3|_{(s,\lambda)=(1,0)},
\end{align}
where $G_{s}$ and $G_{\lambda}$ are the first derivatives of $G(s,\lambda)$ with respect to $s$ and $\lambda$,
and $G_{ss}$, $G_{s\lambda}$ and $G_{\lambda\lambda}$ are the three second derivatives of $G(s,\lambda)$.
Then we can easily evaluate the shot noise and the Fano factor.

When the symmetry between the two flavors of quasiparticles exists
the expression for the Fano factor at zero temperature simplifies:
\begin{align}
\lambda_0'(1)&= 16A/(4+\sum_{k=0}^3 \frac{p_{2k+1}}{p_{2k+4}})\nonumber \\
e^* = & \frac{e\lambda_0'(1)^2}{16}\Bigg\{\frac{1}{4A} \sum_k \frac{p_{2k+1}}{p_{2k+4}^2}-\frac{1}{16A^2}(\sum_k\frac{p_{2k+1}}{p_{2k+4}})^2\nonumber \\
& + \frac{1}{2A^2}\left[1+\frac{1}{4}(\frac{p_3}{p_6}+\frac{p_7}{p_2})(\frac{p_1}{p_4}+\frac{p_5}{p_0})\right]
\Bigg\}\label{fano_sym}
\end{align}
where the identity  $p_{k}+p_{k+4}=2A$ is used to simplify the expressions.
Like the current, the Fano factor $e^*$ is a periodic function of the magnetic flux $\Phi$ with the period $\Phi_0$.
Numerical analysis of Eq.~(\ref{fano_sym}) shows that maximal $e^*$ is $1.4e$, greater than $1.0e$ for Laughlin states but smaller than $3.2e$ for the Pfaffian state.
The maximal Fano factor is achieved at $u=1$. This value of $u$ corresponds to  $\Gamma_1=\Gamma_2$ and $r_1=r_2$. As can be seen from a renormalization group picture, the latter relation between $r_i$
is satisfied at low voltages and temperatures, $eV, T < hv/a$, where $v$ is the velocity of the slowest edge mode and $a$ the distance between the point contacts
along the edges of the interferometer. The values of $\Gamma_k$
can be controlled with gate voltages. Note that $u=1$ is the maximal possible value of $u$.
Indeed, $u>1$ would result in negative probabilities (\ref{eqp})
at some values of the magnetic flux.

An interesting situation emerges in the special limit when only particles of one flavor $a$ or $b$ can tunnel
(i.e., $A^a=0$ or $A^b=0$).
The interferometer can be tuned to that limit with the following approach: one keeps the filling factor $\nu=5/2$ in the gray (green online) region in Fig.~1 and $\nu=0$ in the white region. In addition,
a narrow region with a filling factor $7/3$ is created along the edges. In such situation,
quasiparticles tunnel through the 5/2-liquid between the interfaces of the $\nu=5/2$ and $\nu=7/3$ regions at the tunneling contacts. Since the interface contains only one edge mode, only one type of quasiparticles can tunnel (for a detailed discussion of the interface mode
see Appendix, subsection 1).

We will assume for certainty that $p_k^b=0$. This case is illustrated in Fig.~\ref{fig2} (c).
The calculation of the zero-temperature current and  Fano factor
greatly simplifies in that limit.
One finds expressions resembling the results for the Laughlin states \cite{feldman07}:
\begin{equation}
I=\frac{2e}{\sum_{k=0}^7 1/p_k};
\end{equation}
\begin{equation}
e^*=2e\frac{\sum 1/p_k^2}{[\sum 1/p_k]^2}.
\label{eqasnoise}
\end{equation}
Combining the above equations with Eq.~(\ref{eqp}) one can see
that in the special case of only one type of quasiparticles allowed to tunnel, the current and noise are periodic functions of the magnetic flux
with the ``superconducting'' period $\Phi_0/2$. This is certainly compatible with the Bayers-Yang theorem \cite{BYt}.
The numerator of the fraction in Eq.~(\ref{eqasnoise}) is always smaller or equal to the denominator. They become equal when one of the probabilities
$p_k$ approaches zero.
In that case the Fano factor is maximal and $e^*=2e.$ As is clear from Eq.~(\ref{eqp}), this can happen only at $u=1$. Note that the period and maximal Fano factor are also $\Phi_0/2$ and
$2\times 2e$ in the $K=8$ Abelian state \cite{overbosch08}.

Finally, let us discuss the most general case with nonzero $A^a$ and $A^b$ and no flavor symmetry.
Generally, when the flavor symmetry is absent, $e^*(\Phi)$ is a periodic function with the period $\Phi_0$.
Our numerical results show that the maximal $e^*$ is
achieved
 at $u^a=u^b=1$ for any choice of $A^a$, $A^b$, $\delta^a$ and $\delta^b$.
We calculated the maximal $e^*\equiv e_{\rm max}^*(\gamma,\delta)$ as a function of the tunneling amplitude ratio $\gamma=A^b/A^a$ and phase difference $\delta = \delta^b-\delta^a$. We find that the maximal value of $e^*_{\rm max}=2.3e$ is achieved at $\gamma=0.08$ and $\delta=-0.03$. The dependence of $e^*_{\rm max}$ on $\gamma$ and $\delta$ is shown
 in Fig.~\ref{fig3D}.
We see that at $\gamma=0$, $e^*_{\rm max}$ is $2e$, increases to $2.3e$ at small positive $\gamma$ and drops to $1.4e$ at $\gamma=1$.

In a general case, the current and noise depend on several parameters and the system may not be tuned to the regime with the maximal Fano
factor $2\times 2.3e$. At the same time, such tuning is possible in the flavor-symmetric case, in the case when only one flavor tunnels and in the Pfaffian state. In all those cases one just needs to achieve $u=1$ which corresponds to
$\Gamma_1=\Gamma_2$. Making $\Gamma_{1,2}$ equal is straightforward: one just has to make sure that
the current is the same when only QPC1 or only QPC2 is open. Thus, in the absence of the flavor symmetry, the identification of the 331 states simplifies by
operating the interferometer
in the regime in which only one quasiparticle flavor can tunnel.

\section{Summary}
\label{sec:summary}

We have calculated the current and noise in the Mach-Zehnder interferometer in the 331 state and compared the results with those for the
Pfaffian state. Note that the transport behavior is essentially the same in the Pfaffian and anti-Pfaffian states.
The current dependence on the magnetic flux turns out to be quite similar
for the Pfaffian and 331 states. In both states the $I$-$V$ curves are asymmetric. The states can be unambiguously distinguished with a shot
noise measurement. In the Pfaffian state the maximal Fano factor is $2\times3.2e$. In the 331 state the Fano factor cannot exceed $2\times 2.3e$.
If the flavor symmetry is present than the maximal Fano factor drops to $2\times 1.4e$.
The difference of the predicted maximal Fano factors well exceeds the current
experimental accuracy of 15 percent \cite{heiblum-review}.
An interesting situation emerges, if quasiparticles of
only one flavor can tunnel through the interferometer. In that case the current and noise are periodic functions of the magnetic flux with the period
$\Phi_0/2$. In a general case the period is $\Phi_0$.
We show that it is possible to tune the interferometer
to the regime with the period $\Phi_0/2$ in the 331 state.
\\

\begin{acknowledgments}
We thank B. Rosenow for a useful discussion.
This work was supported by NSF under Grant No. DMR-0544116 and BSF under Grant No. 2006371.
\end{acknowledgments}

\appendix
\section{Multi-component Halperin states}

In this appendix we consider a general multi-component Halperin state \cite{wenzee92} with the filling factor $\nu=2+k/(k+2)$. In the first section of the appendix we describe the states.
In the second section we calculate the current and noise in  the situation in which quasiparticles of one flavor dominate transport through the interferometer.
This is the case of main interest for $k>2$.
In the third section we consider a general situation. Below we ignore the lowest filled Landau level and concentrate on the fractional quantum Hall effect in the second Landau level.
Indeed, the most relevant tunneling operators involve only the fractional edge modes.

\subsection{Quasiparticle statistics and edge modes}

The multi-component Halperin state \cite{wenzee92} with the filling factor $\nu=2+k/(k+2)$ can be described by a $k\times k$ matrix $(K_{\rm mult})_{ij}=1+2\delta_{ij}$. All components of the charge vector ${\bf t}$ equal 1
and the most relevant quasiparticles are described by vectors ${\bf l}_i=(0,\dots,0,1,0,\dots,0)$, where the number 1 stays in position $i$. The charge of the elementary excitations is
$q=e/(k+2)$. The state of the interferometer is described by the numbers $n_1,\dots,n_k$ of the trapped quasiparticles of each of the $k$ types. The statistical phase,
accumulated by a particle of type ${\bf l}_m$ going around the hole in the interferometer, is
\begin{equation}
\label{A0}
\theta=2\pi \sum_p K^{-1}_{mp}n_p=\pi [n_m-n/(k+2)],
\end{equation}
where $n=\sum n_p$.
Hence, the tunneling probability for a particle of type $m$ is
\begin{equation}
\label{A1}
p=A_m [1+u_m\cos(\frac{2\pi \Phi}{(k+2)\Phi_0}+\delta_m+\pi n_m -\frac{\pi n}{k+2})],
\end{equation}
where $A_m$, $u_m$ and $\delta_m$ are real constants.

An alternative description of the same topological state of matter can be formulated in terms of single-component hierarchical states. The starting point is the $\nu=1/3$ Laughlin state.
Condensation of quasiparticle pairs on top of the Laughlin state gives rise to the $\nu=2/(2+2)$ hierarchical state. Condensation of its quasiparticle pairs results in the $\nu=3/(3+2)$ state and so on. The appropriate $K$-matrix $K_{\rm h}$ has the following nonzero elements:
$K_{11}=3$, $K_{nn}=4$ ($k\ge n>1$), $K_{n,n-1}=K_{n-1,n}=-2$. The $K$-matirx $K_{\rm mult}$ expresses via $K_{\rm h}$ as $K_{\rm mult}=WK_{\rm h}W^T$, where $W_{ij}=1$, if $i+j\le (k+1)$, and
$W_{ij}=0$, if $i+j>k+1$. The charge vector reads ${\bf t}=(1,0,\dots,0)$. The $l$-vectors of the most relevant quasiparticle excitations are ${\bf l}=(1,-1,0,\dots,0)$,
$(0,1,-1,0,\dots,0),\dots$, $(0,\dots,0,1,-1)$, $(0,\dots,0,1)$.

In the main part of the article we considered $k=2$ and focused on the flavor-symmetric case. This was justified for $k=2$ for two reasons. First, in the flavor-symmetric
case, the Fabry-P\'{e}rot interferometry cannot distinguish the Pfaffian and 331 states. Second, for the unpolarized 331 state the role of the flavors is played by the electron spin and an approximate symmetry between two spin projections in quantum Hall systems may give rise to the flavor symmetry. At $k\ne 2$ neither reason applies.
Indeed, the Fabry-P\'{e}rot interferometry can distinguish the states with $k\ne 2$ from proposed non-Abelian states \cite{stern09} and $k>2$ different flavors cannot be reduced to different spin projections.
In the absence of the flavor symmetry one expects different tunneling probabilities for different quasiparticle types. As explained below, we expect, in general, that for one quasiparticle type the tunneling probability is much higher than for the other quasiparticles.

To understand why this happens we need to identify edge channels in a $\nu=k/(k+2)$
quantum Hall liquid. For this purpose we need to diagonalize the first term in the edge action
\cite{wen04-book},
$L_k=\frac{1}{4\pi}\int dtdx
[K_{{\rm h},ij}\partial_t\phi_i\partial_x\phi_j-V_{ij}\partial_x\phi_i\partial_x\phi_j]$.
Note that we use the language of hierarhical states.
The diagonalization can be accomplished with the new variables $\Phi_n=(n+2)\phi_n-(n+1)\phi_{n+1}$,
$n=1,\dots,k-1$ and $\Phi_k=(k+2)\phi_k$. The action assumes the form
\begin{equation}
\label{A2}
L_k=\frac{1}{4\pi}\int dt dx [\sum_{n=1}^k K^0_n\partial_t\Phi_n\partial_x\Phi_n-
\tilde V_{ij}
\partial_x\Phi_i\partial_x\Phi_j],
\end{equation}
where $K^0_n=2/[(n+1)(n+2)]$. The operator $\psi_n=\exp(i\Phi_n)$ creates an excitation with one electron charge. The coefficients $K^0_n$ and the charge created by $\psi_n$ are independent of $k$. This allows for an easy description of the interface between the
$\nu=k/(k+2)$ and $\nu=(k-1)/(k+1)$ liquids. Indeed, the action for the interface assumes the following form
\begin{align}
\label{A3}
L=&\frac{1}{4\pi}\int dt dx [\sum_{n=1}^k K_n^0\partial_t\Phi^k_n\partial_x\Phi^k_n-
\sum_{n=1}^{k-1} K_n^0\partial_t\Phi^{k-1}_n\partial_x\Phi^{k-1}_n] \nonumber\\
& + {\rm electrostatic~interaction} + {\rm interchannel~tunneling},
\end{align}
where superscripts $k$ and $k-1$ refer to the different sides of the interface. The minus sign before the second term in the action reflects the opposite propagation directions for the
edge states of the two adjacent liquids. The operators $\exp(i\Phi_n^p)$ create equal charges
for both values of $p=k,k-1$. The prefactors in front of
$\partial_t\Phi^p_n\partial_x\Phi^p_n$ are opposite
for different values of $p$. Hence, according to the criterion
of stability of edge states \cite{haldane,kao}, the modes $\Phi_n^k$
and $\Phi_n^{k-1}$ gap each other out for each $n<k$. Thus, the low-energy interface action reduces
to $L=1/{(4\pi)}\int dt dx [K^0_k\partial_t\Phi^k_k\partial_x\Phi^k_k-V(\partial_x\Phi^k_k)^2]$
and contains a single mode $\Phi_k^k$.

Let us now consider a system whose filling factor changes in a step-wise manner: as one moves across the edge of the quantum Hall bar, the filling factor first changes from 0 to $1/3$, then to $2/(2+2)$, then to $3/(3+2)$ etc.
The filling factor in the innermost part of the sample is $k/(k+2)$. The edge action can still be written in the form (\ref{A2}). Different modes $\Phi_n$ correspond to
spatially separated interfaces between consecutive regions with different filling factors.
A quantum point contact in such systems brings
close to each other two innermost edge channels, corresponding to the mode
$\Phi_k$. Clearly, any tunneling process can involve only those two interfaces.
Hence, only one type of the quasiparticles with the $l$-vector $(0,\dots,0,1)$ is allowed to tunnel. By shrinking the regions with intermediate filling factors $0<\nu<k/(k+2)$ one can deform the above system into a quantum Hall bar with a single interface between the $\nu=0$
and $\nu=k/(k+2)$ regions. Still, it is clear that the innermost edge channel corresponds to the field $\Phi_k$. The tunneling probability rapidly decreases as the path, traveled by the
tunneling particle, grows. Hence, the maximal tunneling probability corresponds to
quasiparticles of type $(0,\dots,0,1)$ and it is reasonable to neglect all other tunneling processes.

Certainly, this is an approximation. In general, the tunneling probabilities (\ref{A1})
depend on the parameters $A_m$, $u_m$  and $\delta_m$. Note that one phase $\delta_m$
can be excluded by absorbing it into the
Aharonov-Bohm phase due to the magnetic flux $\Phi$. Two more parameters can be excluded by tuning gate voltages at the point contacts. Still $3(k-1)$ fitting parameters remain. An expression with a large number of fitting parameters is of limited use. Fortunately, the model with only one type of tunneling quasiparticles becomes exact, if the edges of the Mach-Zehnder interferometer
correspond to the interface between $\nu=k/(k+2)$ and $\nu=(k-1)/(k+1)$ liquids.
This can be accomplished if in addition to the gray (green online) region with the filling factor
$2+k/(k+2)$ in Fig.~1 and the white region with the filling factor $0$ one creates a strip with the filling factor $2+(k-1)/(k+1)$ along the edges.
As shown in the next section, it is possible to derive an expression for the Fano factor without any fitting parameters in that case. This situation will be our main focus below.

\subsection{The case of only one flavor allowed to tunnel}

In this section we will omit the index $m$ in Eq.~(\ref{A1}) since it can assume only one value.
We will also set $\delta_m=0$ since it can be absorbed in the Aharonov-Bohm phase.
We will focus on the limit of low temperatures so that quasiparticles
only tunnel from the edge
with the higher potential to the edge with the lower potential.
The tunneling probability depends only on the number $n$ of the
trapped quasiparticles and reads
\begin{equation}
\label{A4}
p_n=A [1+u\cos(\frac{2\pi \Phi}{(k+2)\Phi_0}+\frac{\pi n(k+1)}{k+2})],
\end{equation}
The number of different probabilities $p_n$ depends on the parity of $k$. For odd
$k$, $p_n=p_{n+k+2}$ and hence we can define $n$ modulo $k+2$. For even $k$, $p_n=
p_{n+2(k+2)}\ne p_{n+k+2}$ and $n$ should be defined modulo $2(k+2)$. Hence, the number of different values of $p_n$ equals
$n_{\rm max}=(k+2)[3+(-1)^{k+2}]/2$.
In what follows we will see a considerable difference between even and odd $k$.

It will be convenient to use the following parametrization of Eq.~(\ref{A4})
\begin{equation}
\label{A5}
p_n=B[v\exp(-i\phi_0)+\exp(in\theta)][v\exp(i\phi_0)+\exp(-in\theta)],
\end{equation}
where $B=A[1-\sqrt{1-u^2}]/2$, $v=[1+\sqrt{1-u^2}]/u$, $\phi_0=2\pi\Phi/[(k+2)\Phi_0]$ and $\theta=\pi(k+1)/(k+2)$.
All exponents $\exp(in\theta)$ are $n_{\rm max}$th roots of 1.

\subsubsection{Current}

We now use Eq.~(\ref{A5}) for the calculation of the current. The current $I=\lim_{t\rightarrow\infty} Nq/t$,
where $q=e/(k+2)$ is a quasiparticle charge and $N$ is the large number of tunneling events
over a long time period $t$. It is easy to find the average time $t_n$ between the tunneling events which change the number of the trapped quasiparticles from $n-1$ to $n$ and from $n$ to $n+1$: $t_n=1/p_n$.
Taking into account that there are only $n_{\rm max}$ different
values of $p_n$, one finds the average time of $n_{\rm max}$ consecutive tunneling events:
\begin{equation}
\label{A6}
\bar t=n_{\rm max}\lim_{t\rightarrow\infty}t/N=\sum_{n=1}^{n_{\rm max}}1/p_n.
\end{equation}
Hence,
\begin{equation}
\label{A7}
I=\frac{n_{\rm max}q}{\sum \frac{1}{p_n}}.
\end{equation}

We next evaluate the sum $\bar t=\sum 1/p_n$. Similar sums were evaluated numerically in the main part of the article. For an arbitrary $k$ we need an analytic expression. Two different expressions hold for even and odd $k$. We will only discuss the details for even $k$ as the case of odd $k$ can be considered in a similar way.

Using the parametrization (\ref{A5}) one gets
\begin{equation}
\label{A8}
\bar t = \frac{\sum_n |Q_n(v\exp(-i\phi_0))|^2}{B|P(v\exp(-i\phi_0))|^2},
\end{equation}
where $P(z)=\prod_l(z+\exp(il\theta))$ and $Q_n(z)=P(z)/(z+\exp(in\theta))$.

$P(z)$ can be found from the basic theorem of algebra. Indeed, the roots of that polynomial
are known and equal $-\exp(il\theta)$. There is only one such polynomial of power
$n_{\rm max}=2(k+2)$ with the coefficient 1 before $z^{n_{\rm max}}$. This polynomial is
$P(z)=z^{n_{\rm max}}-1$. From this one gets
$Q_n(z)=\sum_{l=0}^{n_{{\rm max}-1}}z^l(-1)^{l+1}\exp(-i\theta n[l+1])$.
The sum $\sum_n|Q_n|^2$ can now be calculated by first performing the summation over $n$ and reduces to $\sum_n|Q_n|^2=n_{\rm max}(v^{2n_{\rm max}}-1)/(v^2-1)$. Finally, for even $k$
\begin{equation}
\label{A9}
\bar t_e=\frac{n_{\rm max}(v^{2n_{\rm max}}-1)}{B(v^2-1)[v^{2n_{\rm max}}-2v^{n_{\rm max}}
\cos n_{\rm max}\phi+1]};
\end{equation}
\begin{equation}
\label{A10}
I_e=
\frac{eB(v^2-1)}{k+2}
\frac{v^{4k+8}-2v^{2k+4}
\cos\frac{4\pi\Phi}{\Phi_0}+1}
{v^{4k+8}-1}.
\end{equation}
For odd $k$, a similar calculation yields
\begin{equation}
\label{A11}
I_o=\frac{eB(v^2-1)}{k+2}\frac{v^{2k+4}+2v^{k+2}
\cos\frac{2\pi\Phi}{\Phi_0}+1}{v^{2k+4}-1}.
\end{equation}

The expressions look similar for even and odd $k$ but there is a significant difference between them. Indeed, for odd $k$, the current is a periodic function of the magnetic flux with the period $\Phi_0$. For even $k$ the period is two times shorter.
Such superconducting periodicity reflects Cooper pairing of composite fermions.

\subsubsection{Noise}

The noise is defined as $S=2\lim_{t\rightarrow\infty}(\overline{Q^2}-{\bar Q}^2)/t$,
where $Q$ is the total charge transmitted during the time period $t$. Let us
set $t=m\bar t$, where $m$ is a large integer. Such choice of $t$ corresponds on average to $N=mn_{\rm max}$ tunneling events. The total time required for $N$ tunneling events can be
expressed as $\tau=m\bar t+\delta \tau$, where $\delta \tau$ is a fluctuation. The total charge transferred through the interferometer after $N$ events is exactly $Nq$. In a good approximation, the charge transferred over the time $t=m\bar t$ is then $Q=Nq-I\delta \tau$,
where $I$ is the average current (\ref{A10},\ref{A11}). Substituting this expression in the definition of the noise one gets $S=2I^2\overline{\delta \tau^2}/t$. The calculation of
$\overline{\delta\tau^2}$ is straightforward. One finds: $\overline{\delta\tau^2}=m\delta t^2$,
where $\delta t^2=\sum_{n=1}^{n_{\rm max}}1/p_n^2$. Finally, the Fano factor
\begin{equation}
\label{A12}
e^*=\frac{S}{2I}=n_{\rm max} q\frac{\sum 1/p_n^2}{(\sum 1/p_n)^2}.
\end{equation}

Some restrictions on $e^*$ are evident from elementary inequalities.
Obviously, $e^*\le n_{\rm max}q$.
The inequality
of quadratic and arithmetic means also implies that $e^*\ge q$.
The first inequality sets different upper limits on the effective charge $e^*$ for even and odd
$k$. For odd $k$, $e^*<e$. For even $k$, $e^*<2e$.
This difference agrees with the difference of the magnetic field dependences of the current,
discussed above.
Both upper limits can be reached as
we will see below.

As in the calculation of the current, we focus on the case of even $k$ and only give the final answer for odd $k$. We need to find $\delta t^2$. One can easily see that
in the notation of the previous subsection
\begin{align}
\label{A13}
\delta t^2=&\frac{\sum_{n=0}^{n_{\rm max}-1}|Q_n(v\exp(-i\phi_0))|^4}{B^2|P(v\exp(-i\phi_0))|^4}\nonumber\\
=&\frac{n_{\rm max}}{B^2|P|^4}\sum_{l,m,p,q=0}^{n_{\rm max}-1}
(-v)^{l+m+p+q}\exp(i\phi_0[p+q-l-m]) \nonumber\\ &
\times\sum_s\delta(p+q-l-m-n_{\rm max}s),
\end{align}
where $s$ is an integer and the discrete delta function $\delta(0)=1$, $\delta(a\ne 0)=0$.
The remaining summation is tedious but straightforward.  For even $k$ we obtain
\begin{eqnarray}
\label{A14}
e_e^*=\frac{e}{k+2}(\frac{w-1}{w^{2k+4}-1})^2 [
\frac{d}{dw}w\frac{d}{dw}\frac{w^{2k+5}-1}{w-1}+& & \nonumber\\
(4k+7-w\frac{d}{dw})^2\frac{w^{4k+7}-w^{2k+4}}{w-1}+ & & \nonumber\\
2w^{k+2}\cos(\frac{4\pi \Phi}{\Phi_0})(2k+3-w\frac{d}{dw})\frac{d}{dw}\frac{w^{2k+4}-1}{w-1}], & &
\end{eqnarray}
where $w=[1+\sqrt{1-u^2}]^2/u^2$.

A similar calculation for odd $k$ yields
\begin{eqnarray}
\label{A15}
e_o^*=\frac{e}{k+2}(\frac{w-1}{w^{k+2}-1})^2 [
\frac{d}{dw}w\frac{d}{dw}\frac{w^{k+3}-1}{w-1}+& & \nonumber\\
(2k+3-w\frac{d}{dw})^2\frac{w^{2k+3}-w^{k+2}}{w-1}- & & \nonumber\\
2w^{(k+2)/2}\cos(\frac{2\pi \Phi}{\Phi_0})(k+1-w\frac{d}{dw})\frac{d}{dw}\frac{w^{k+2}-1}{w-1} ], & &
\end{eqnarray}
The expressions are rather similar for even and odd $k$ but their periodicity as a function of the magnetic flux
is  different just like for the current.

For a general $u$, the above expressions are complicated.
They simplify in an important limit case considered below.
By tuning
the gate voltages at the tunneling contacts it is always possible to make equal the tunneling amplitudes at the two contacts. The desired situation can be achieved by selecting such gate voltages
that the total currents would be the same when only one contact is open no matter which one it is.
As discussed in section V, this corresponds to $u=1$ at a sufficiently low temperature and voltage bias.
The expressions for the Fano factor greatly simplify in that limit. For even $k$,
\begin{equation}
\label{A16}
e_e^*=e\frac{8(k+2)^2+1+[4(k+2)^2-1]\cos\frac{4\pi\Phi}{\Phi_0}}{6(k+2)^2}.
\end{equation}
For odd $k$,
\begin{equation}
\label{A17}
e_o^*=e\frac{2(k+2)^2+1-[(k+2)^2-1]\cos\frac{2\pi\Phi}{\Phi_0}}{3(k+2)^2}.
\end{equation}
As a function of the flux, $e_e^*$ oscillates between $e[\frac{2}{3}+\frac{1}{3(k+2)^2}]$ and $2e$. $e_o^*$ oscillates between $e[\frac{1}{3}+\frac{2}{3(k+2)^2}]$ and $e$. The maximal values of the Fano factor are thus $2\times e$ at odd $k$ and $2\times 2e$ at even $k$. The minimal values
of the Fano factor uniquely identify states with different $k$.

\subsection{General case}

Here we address the situation when all quasiparticle flavors are allowed to tunnel. As discussed above, this situation is less interesting than the case of just one flavor allowed to tunnel. Indeed, the expressions for the current and noise depend on numerous fitting parameters and thus are less useful than the results (\ref{A16},\ref{A17}). Besides, in a general case it would be difficult to tune the system to reach its maximal possible
Fano factor. Instead, there is an interval of relevant Fano factors and this makes the identification of a state more difficult. Nevertheless, we discuss below how one can calculate the transport properties for a general $k$ when all flavors are allowed to tunnel.

The current can be found from the system of equations of the form Eq.~(\ref{eq:current},\ref{eq:steady}).
$f_l$ in these equations are the probabilities to find the trapped topological charge $l$ in the interferometer. Thus, it is important to understand how many
different values of the topological charge are possible.

Different topological charges correspond to different $n_p$ in tunneling probabilities (\ref{A1}). Different
$n_p$ may however describe the same topological state. This happens, if all tunneling
probabilities (\ref{A1}) are the same for a certain set of $n_p$ and another set of $n_p+\Delta_p$. The probabilities are equal, if for each $m$
\begin{equation}
\label{B1}
\pi\Delta_m-\frac{\pi\sum_p\Delta_p}{k+2}=2\pi r_m,
\end{equation}
where $r_m$ is an integer. One finds from (\ref{B1}) that
$\Delta_m=2r_m+\sum_p\Delta_p/(k+2)$. Adding up such equations for all $m$, one finds
that $\sum_p\Delta_p=(k+2)\sum_p r_p$ and hence
\begin{equation}
\label{B2}
(\Delta_1,\dots,\Delta_k)=\sum_{m=1}^k r_m {\bf d}_m,
\end{equation}
where ${\bf d}_m=(1,\dots,1,3,1,\dots,1)$ and the number 3 stays in position $m$.
Eq.~(\ref{B2}) means that adding or subtracting a vector ${\bf d}_m$ from
the $l$-vector ${\bf n}=(n_1,\dots,n_k)$ does not change the topological charge. ${\bf d}_m$ can be understood as $l$-vectors of electrons.

If we choose $r_k=1$ and $r_p=-1$ for one $p<k$, then adding the $l$-vector (\ref{B2})
to ${\bf n}$ increases $n_k\rightarrow n_k+2$ and decreases $n_p\rightarrow n_p-2$.
Setting $r_k=-1$ and $r_p=1$
corresponds to the operation $n_k\rightarrow n_k-2$, $n_p\rightarrow n_p+2$.
This operation involves adding to ${\bf n}$ one vector ${\bf d}_m$ and
subtracting another vector ${\bf d}_n$.
Performing several such operations, one can always reduce all $n_p$ with $p<k$ to zeroes and ones.
Choosing $r_k=k+1$ and $r_p=-1$ for all $p<k$ corresponds to the operation
$n_k\rightarrow n_k+2(k+2)$, $n_p\rightarrow n_p$.
The operation involves $2(k+1)$ additions and subtractions of ${\bf d}_m$'s.
Setting $r_k=-(k+1)$ and $r_p=1$ for all $p<k$
corresponds to $n_k\rightarrow n_k-2(k+2)$, $n_p\rightarrow n_p$. Hence, by adding and subtracting vectors ${\bf d}_m$ one can always reduce $l$-vectors to topologically
equivalent vectors of the form $(s_1,\dots,s_k)$, where $s_p=0,1$ at $p<k$ and $ 0 \le s_k<2(k+2)$. We will call such vectors allowable. The total number of the allowable vectors is
$2^k(k+2)$.

We have established that any $l$-vector can be reduced to the allowable form by an even number
of additions and subtractions of vectors ${\bf d}_m$. Moreover, one can get exactly one allowable vector
${\bf N}_e({\bf n})$ from each ${\bf n}$ by an even number of additions and subtractions. Indeed,
an even number of additions and subtractions does not change the parity of each component of the $l$-vector. This fixes $s_p$, $p<k$. The residue $(\sum n_m)~{\rm mod}~2(k+2)$ is also invariant with respect to an even number of subtractions and additions. This fixes $s_k$.
As a consequence, no other allowable vectors can be obtained from an allowable vector by an even number of additions and subtractions.

Next, note that ${\bf N}_o({\bf n})={\bf N}_e({\bf n}+{\bf d}_1)$ is the only allowable vector
that can be obtained from ${\bf n}$ by an odd number of additions and subtractions. Indeed,
if another allowable vector ${\bf M}_o({\bf n})$ could be obtained from ${\bf n}$ by an odd number of additions and subtractions then ${\bf N}_o({\bf n})$ and ${\bf M}_o(\bf n)$ could be obtained from each other by an even number of additions and subtractions: one first gets ${\bf n}$ from ${\bf M}_o(\bf n)$ and then ${\bf N}_o({\bf n})$ from ${\bf n}$. We have already proved that this is impossible.

Thus, exactly one allowable vector ${\bf N}_e({\bf n})$ can be obtained from ${\bf n}$ by an even number of additions and subtractions and exactly one allowable vector
${\bf N}_o({\bf n})$ can be obtained by an odd
number of additions and subtractions. Note that ${\bf N}_e({\bf n})\ne {\bf N}_o({\bf n})$
since the parity of each component $s_p$ changes after each addition or subtraction and hence
the parities of the components of ${\bf N}_e({\bf n})$ and ${\bf N}_o({\bf n})$ are opposite.
Note also that ${\bf N}_e({\bf n})={\bf N}_o( {\bf N}_o({\bf n}))$
and ${\bf N}_o({\bf n})={\bf N}_o( {\bf N}_e({\bf n}))$. Thus, the set of allowable vectors
consists of $2^{k-1}(k+2)$ pairs of the form $({\bf V},{\bf N}_o({\bf V}))$. Different pairs have no common elements. $l$-vectors from different pairs cannot be obtained from each other
by any number of additions and subtractions. If an $l$-vector can be reduced to the allowable
vector ${\bf V}$ by an even number of additions and subtractions it can also be reduced
to ${\bf N}_o({\bf V})$ by an odd number of additions and subtractions. Let us now select
one arbitrary vector from each pair of allowable vectors $({\bf V},{\bf N}_o({\bf V}))$.
We get a set $\gimel$ of $2^{k-1}(k+2)$ allowable vectors. An arbitrary $l$-vector $(n_1,\dots,n_k)$ is topologically equivalent to one of the vectors in $\gimel$. Hence,
there are $2^{k-1}(k+2)$ different topological charges.

Introducing a probability distribution $f_l$ for
different trapped topological charges $l$, one can compute the current from the system
of equations (\ref{eq:current},\ref{eq:steady}). Even for $k=2$ the result is complicated
and in the main part of the article we showed it only in the graphical form (Figs. 3, 4) except for the flavor-symmetric case when it simplifies considerably.

A simplification is also possible in the flavor-symmetric case for a general  $k$.
In that case
one can remove the index $m$ from the constants $A_m$, $u_m$ and $\delta_m$ since they do not depend on the flavor.
 Let us
set $n=(\sum n_p)~ {\rm mod}~ 2(k+2)$ and $S=\sum (n_p ~{\rm mod}~ 2)$, i.e., let $n$ denote the total number of trapped quasiparticles modulo $2(k+2)$
 and $S$ show how many $n_p$'s are odd.
Note that $n$ and $S$ have the same parity. The number of possible pairs $(n,S)$ equals
$C=(k+1)(k+2)$.
Let us introduce a distribution function $f_{n,S}$.
A simplification of the steady state equations results
from the fact that one can write a closed set of equations for $f_{n,S}$:

\begin{eqnarray}
\label{A18}
0=\frac{df_{n,S}}{dt}=-\{A[1-u\cos(\frac{2\pi\Phi}{(k+2)\Phi_0}+\delta-\frac{\pi n}{k+2})]S
& & \nonumber\\
+A[1+u\cos(\frac{2\pi\Phi}{(k+2)\Phi_0}+\delta-\frac{\pi n}{k+2})](k-S)\}\times f_{n,S} & &
\nonumber\\
+f_{n-1,S-1}(k-S+1)A[1+u\cos(\frac{2\pi\Phi}{(k+2)\Phi_0}+\delta-\frac{\pi (n-1)}{k+2})] & &
\nonumber\\ +
f_{n-1,S+1}(S+1)A[1-u\cos(\frac{2\pi\Phi}{(k+2)\Phi_0}+\delta-\frac{\pi (n-1)}{k+2})], & &
\end{eqnarray}
where the notation implies that $f_{n,S}=0$ for $S<0$ and $S>k$.
The current can then be expressed as
\begin{eqnarray}
\label{A19}
I=\sum f_{n,S}\{A[1-u\cos(\frac{2\pi\Phi}{(k+2)\Phi_0}+\delta-\frac{\pi n}{k+2})]S
& & \nonumber\\
+A[1+u\cos(\frac{2\pi\Phi}{(k+2)\Phi_0}+\delta-\frac{\pi n}{k+2})](k-S)\} & &
\end{eqnarray}

System (\ref{A18}) contains $C=(k+1)(k+2)$ equations, much fewer than
$(k+2)2^{k-1}$ in a general case. Moreover, one can reduce the total number of equations to no more than $k/2+1$. Indeed, Eq. (\ref{A18}) allows one to express $f_{n,S}$ via $f_{n-1,S\pm 1}$. $f_{n-1,S\pm 1}$ can be expressed via $f_{n-2,S'}$, where $S'=S-2,S,S+2$, etc. After $2(k+2)$ steps, one expresses $f_{n,S}$ via the values of the distribution function $f_{n-2(k+2),S}$, where $S=0,\dots, k$. Since $n$ is defined modulo $2(k+2)$, this means that a closed system can be obtained for $k+1$ variables $f_{n,S}$ for any fixed $n$.
At least $k/2$ of those variables are zeroes since $n$ and $S$ have the same parity
for any nonzero $f_{n,S}$. This leaves no more than $k/2+1$ equations.
 After the system is solved, one can immediately compute $f_{n+1,S}$ from $f_{n,S}$ with Eq. (\ref{A18}),
then express $f_{n+2,S}$ via $f_{n+1,S}$ and so on.

\end{document}